\documentclass[aps,prd,preprint,floatfix,nofootinbib,showpacs,superscriptaddress]{revtex4-1}
\usepackage{dcolumn}
\usepackage{bm}
\usepackage{graphicx,xcolor}
\usepackage{amssymb,amsmath}
\usepackage{multirow}
\usepackage{units}
\usepackage{color,url}
\usepackage{slashed}
\usepackage{tabu}
\usepackage{array}
\usepackage[colorlinks=true,urlcolor=blue,anchorcolor=blue
,citecolor=blue,filecolor=blue,linkcolor=blue,menucolor=blue
,linktocpage=true,pdfproducer=medialab,pdfa=true]{hyperref}
\usepackage{colordvi}
\def\beqa{\begin{eqnarray}}
\def\eeqa{\end{eqnarray}}

\newcommand{\gev}{\ensuremath{\,\mathrm{GeV}}}

\newcommand{\sv}{\ensuremath{\langle\sigma v\rangle}}
\newcommand{\gtwo}{\ensuremath{\Delta a_\mu}}

\definecolor{darkgreen}{HTML}{2a9000}

\begin{document}

\title{Shedding light on dark matter with recent muon $(g-2)$ and Higgs exotic decay measurements}
\def\slash#1{#1\!\!/}

\renewcommand{\thefootnote}{\arabic{footnote}}

\author{Chih-Ting Lu}
\email{timluyu@kias.re.kr}
\affiliation{School of Physics, KIAS,\\85 Hoegiro, Dongdaemun-gu, Seoul 02455, Republic of Korea}

\author{Raymundo Ramos}
\email{raramos@gate.sinica.edu.tw}
\affiliation{Institute of Physics, Academia Sinica,\\Nangang, Taipei 11529, Taiwan}

\author{Yue-Lin Sming Tsai}
\email{smingtsai@pmo.ac.cn}
\affiliation{Key Laboratory of Dark Matter and Space Astronomy, Purple Mountain Observatory,
    Chinese Academy of Sciences, Nanjing 210008, China}

\date{\today}

\begin{abstract} 

Recently, we have witnessed two hints of physics beyond the standard model:
A 3.3$\sigma$ local excess ($M_{A_0} = 52$~GeV) in the search for $H_0\to A_0 A_0\to b\overline{b}\mu^{+}\mu^{-}$ and a 4.2$\sigma$ deviation from the SM prediction in the $(g-2)_\mu$
measurement.
The first excess was found by the ATLAS Collaboration using 139~fb$^{-1}$ data
at $\sqrt{s}=13$~TeV\@.
The second deviation is a combination of the results from the
Brookhaven E821 and the recently reported Fermilab E989 experiment.
We attempt to explain these deviations in terms of
a renormalizable simplified dark matter model.
Inspired by the null signal result from dark matter (DM) direct
detection, we interpret the possible new particle, $A_0$, as a pseudoscalar
mediator connecting DM and the standard model. 
On the other hand, a new vector-like muon lepton can explain 
these two excesses at the same time while contributing to the DM phenomenology.

\end{abstract}

\maketitle

\section{Introduction}
The existence of dark matter (DM) is now well established
by old and new astrophysical and cosmological evidence.   
Conversely, its particle properties remain unclear,  
in particular, the way to incorporate DM and its interactions into
the standard model (SM) of particle physics. 
In the case of DM-quark interaction, 
current DM direct detection experiments, such as XENON1T,
have not observed any DM-nuclei
scattering evidence~\cite{Aprile:2018dbl}.     
Furthermore, the Muon $g-2$ Collaboration at Fermilab has recently reported an
eye-catching measurement of
the anomalous magnetic dipole moment of $\mu^{\pm}$ 
with a $3.3\sigma$ deviation from the SM prediction
achieving a combined experimental average of
$\Delta a_{\mu}=(2.51\pm 0.59)\times 10^{-9}$~\cite{Abi:2021gix}.
This confirms a long standing tension between SM and experimental data that
was previously reported by the E821 experiment at Brookhaven National Laboratory,
inspiring several beyond the SM (BSM) extensions (see Ref.~\cite{Lindner:2016bgg} for a review).
The new average is consistent with a $4.2\sigma$ deviation from the SM prediction
strongly motivating
new proposals~\cite{Yin:2021yqy,Chiang:2021pma,Lee:2021jdr,Crivellin:2021rbq,Endo:2021zal,Iwamoto:2021aaf,Han:2021gfu,Arcadi:2021cwg,Criado:2021qpd,Zhu:2021vlz,Gu:2021mjd,Wang:2021fkn,VanBeekveld:2021tgn,Nomura:2021oeu,Anselmi:2021chp,Yin:2021mls,Wang:2021bcx,Buen-Abad:2021fwq,Das:2021zea,Abdughani:2021pdc,Chen:2021jok,Ge:2021cjz,Cadeddu:2021dqx,Brdar:2021pla,Cao:2021tuh,Chakraborti:2021dli,Ibe:2021cvf,Cox:2021gqq,Babu:2021jnu,Han:2021ify,Heinemeyer:2021zpc,Calibbi:2021qto,Amaral:2021rzw,Bai:2021bau,Baum:2021qzx,
Li:2021poy,Zu:2021odn,Keung:2021rps,Ferreira:2021gke,Zhang:2021gun,Ahmed:2021htr,Zhou:2021cfu,Yang:2021duj,Athron:2021iuf,Chen:2021vzk,Chun:2021dwx,Escribano:2021css,Aboubrahim:2021rwz,Bhattacharya:2021ggm}.
Furthermore, the ATLAS collaboration has reported a search for
the Higgs boson exotic decay channel $H_0\to A_0 A_0\to b\overline{b}\mu^{+}\mu^{-}$,
using $139$~fb$^{-1}$ data at $\sqrt{s}=13$~TeV~\cite{ATLAS:2021ypo}. 
They have taken the muon pair as a trigger to search for the narrow dimuon
resonance of a light spin-0 particle $A_0$. 
A deviation from the SM background with a local (global) significance of $3.3\sigma$
(1.7$\sigma$) 
is reported at $M_{A_0} = 52$~GeV with the branching fraction given by
${\rm BR}(H_0\rightarrow A_0 A_0\rightarrow b\overline{b}\mu^{+}\mu^{-})\sim 3.5\times 10^{-4}$. 
Therefore, the null observation from XENON1T and the two possible BSM observations 
($\Delta a_{\mu}$ and $H_0\to A_0 A_0\to b\overline{b}\mu^{+}\mu^{-}$)
can be interpreted as hints that 
the new spin-0 particle $A_0$ can serve as a mediator to connect DM with SM, especially for its non-trivial coupling with muons. 

One can simply consider the additional spin-0 particle $A_0$ coupling to SM fermion pairs 
via the mixing with SM-like Higgs boson only or like 
the scalar/pseudoscalar in the conventional two Higgs doublet models 
(Type-I, II, X, and Y 2HDM~\cite{Branco:2011iw}, 2HDM+S~\cite{Sabatta:2019nfg}) 
with $M_{A_0} = 52$~GeV.  
However, their predicted signal strength for ${\rm BR}(H_0\rightarrow A_0 A_0\rightarrow b\overline{b}\mu^{+}\mu^{-})$ 
cannot reach the measured $3.5\times 10^{-4}$
when including all other experimental constraints~\cite{Curtin:2013fra,ATLAS:2020qdt}. 
We go beyond the models mentioned above and involve a new vector-like muon lepton (VLML) 
not only to enhance ${\rm BR}(A_0\rightarrow\mu^{+}\mu^{-})$
but also to contribute to $\Delta a_{\mu}$~\cite{Hiller:2019mou},
thus, explaining both excesses. 
Motivated by these observations and from theoretical considerations, 
we propose a renormalizable simplified DM model based on 
extending the SM with three SM singlet fields:
a Dirac DM, a VLML, and a pseudoscalar mediator.
One important advantage of adopting a pseudoscalar mediator here is that 
the elastic cross section of DM-nuclei collision is suppressed by 
the small recoil energy making easy to escape the current XENON1T stringent limit.

The paper is organized as follows. First, we briefly discuss the model setup in Section~\ref{sec:model}. 
Next, we consider relevant constraints for this model used in our likelihood functions in Sec.~\ref{sec:constraint}. In Sec.~\ref{sec:result}, we present our numerical analysis and the $2\sigma$ allowed regions. Finally, we conclude in Sec.~\ref{sec:conclusion}.

\section{Renormalizable simplified dark matter model}
\label{sec:model} 
In this section, we show our model configuration.  
We consider a SM singlet Dirac fermion $\chi$ as a DM candidate. 
A new pseudoscalar mediator $A$ is also introduced 
to explain the null signal result
in DM direct detection and possible excess in Higgs exotic decay.  
Additionally, we introduce a VLML, $\psi^{\pm}$,
that will contribute to the $(g-2)_{\mu}$ excess.  
Thus, the renormalizable Lagrangian for this simplified DM model can be written as   
\begin{align}
{\cal L} = {}& 
{\cal L}_{SM} + \overline{\chi}(i\slashed{\partial}-M_{\chi}-ig_{\chi}A\gamma_5)\chi +\frac{1}{2}\partial_{\mu}A\partial^{\mu}A -\frac{1}{2}m^2_A A^2 
\nonumber  \\
&  -(\mu_A A +\lambda_{HA}A^2)(H^{\dagger}H -\frac{v^2}{2}) -\frac{\mu^{\prime}_A}{3!}A^3
-\frac{\lambda_A}{4!}A^4 
\nonumber  \\
& +\left[ -\kappa\overline{L}_{\mu}H\psi_R + i\kappa^{\prime}\overline{\mu}_R A\psi_L
-iy\overline{\psi}_L A\psi_R +M_{\psi}\overline{\psi}_L \psi_R +\text{H.c.}\right]. 
\label{eq:lagrangian}
\end{align} 
where $H$ is the SM scalar SU(2)$_L$ doublet
and $L_{\mu}$ is the second generation left-handed lepton doublet. 
Note that the dimension-3 terms with $\mu_A$ and $\mu^{\prime}_A$
break the parity~\cite{Baek:2017vzd}.
The tadpole for $A$ is removed and $\langle A\rangle = 0$ is assumed in this model. 

After electroweak symmetry breaking (EWSB), the pseudoscalar $A$
and the SM Higgs boson $h$ mix with each other via the $\mu_A$ term.
The relation between the mass eigenstates, $H_0$ and $A_0$,
and the interaction states, $h$ and $A$, is
\begin{equation}
 \left( \begin{array}{c}
                H_0 \\ 
                A_0 \end{array} \right ) 
= 
 \left( \begin{array}{cc} 
            \cos\alpha & \sin\alpha \\
            - \sin\alpha & \cos\alpha \end{array} \right )\,
 \left( \begin{array}{c}
                h \\ 
                A \end{array} \right ) 
\end{equation}
where the mixing angle, $\alpha$, is given by
\begin{equation}
\sin 2\alpha = \frac{2\mu_A v}{M^2_{H_0}-M^2_{A_0}} 
\end{equation} 
with $v\sim 246$~GeV the vacuum expectation value.
We assign $H_0$ as the SM-like Higgs boson with $M_{H_0}=125$~GeV
and $A_0$ with $M_{A_0}=52$~GeV to explain the ATLAS Higgs boson exotic decay excess~\cite{ATLAS:2021ypo}.
In the next section, we will see that the LHC Higgs boson measurements 
can put a strong upper limit on $\sin 2\alpha$ in this model.
Similarly, the VLML and muon will mix together after EWSB~\cite{Hiller:2019mou}. 
According to $Z\rightarrow l^{+}l^{-}$ precision measurements~\cite{Tanabashi:2018oca}, 
an upper limit for the mixing between the left-handed muon and VLML is set at
\begin{equation}
\frac{\kappa v}{\sqrt{2}M_{\psi}} < {\cal O}(10^{-2}). 
\label{Eq:VLML_mix}
\end{equation}
where $M_{\psi}\sim v$ implies $\kappa\lesssim {\cal O}(10^{-2})$.
In contrast, $\kappa^{\prime}$ doesn't suffer from this constraint and its value can be larger. 

The Lagrangian to describe the interactions between the SM sector and DM sector 
via $H_0$ and $A_0$ portal can be written as 
\begin{align}
{\cal L}^{(H_0, A_0)}_{int} = {}& 
-ig_{\chi}\left(H_0 s_\alpha + A_0 c_\alpha\right)\overline{\chi}\gamma_5\chi -\left(H_0 c_\alpha - A_0 s_\alpha\right)
\left[ \sum_{f\neq\mu}\frac{m_f}{v}\overline{f}f -\sum_{V=Z,W^\pm}\frac{\delta_V m^2_V}{v}V_{\mu}V^{\mu}\right]
\nonumber  \\ & 
-\left(\frac{m_{\mu}}{v}c_\alpha + i g_A\gamma_5 s_\alpha \right)H_0\overline{\mu}\mu
+\left(\frac{m_{\mu}}{v}s_\alpha - i g_A\gamma_5  c_\alpha \right)A_0\overline{\mu}\mu 
\label{Eq:HA-int}
\end{align}
where $s_\alpha =\sin\alpha$, $c_\alpha = \cos\alpha$ and in the first-order approximation of $\kappa$,
\begin{equation}
g_A = \frac{\kappa^{\prime}\kappa}{\sqrt{2}}\frac{v}{M_{\psi}} 
\label{Eq:gA}
\end{equation} 
and $\delta_V = 1(2)$ for $V=Z$($W^{\pm}$).
Note that, excepting muon pair, there are only axial (scalar) couplings
for $H_0$ and $A_0$ with DM pair (SM fermion pairs).
The term with $g_A$ is important to enhance ${\rm BR}(A_0\rightarrow\mu^{+}\mu^{-})$.   
Additionally, the CP-violation effect in the muon Yukawa interactions is a feature of this model.
For the first order of $\kappa$ approximation, 
the interactions of VLML can be expanded as 
\begin{align}
{\cal L}^{\psi}_{int} = &
-e\overline{\psi}\gamma^{\mu}\psi A_{\mu}+\frac{g}{c_W}\overline{\psi}\gamma^{\mu}\psi Z_{\mu} 
-iy\overline{\psi}(H_0 s_\alpha + A_0 c_\alpha)\gamma^5\psi 
\nonumber  \\ &
+ \bigg[ -\frac{\kappa}{\sqrt{2}}(H_0 c_\alpha - A_0 s_\alpha)\overline{\mu}_L\psi_R 
+i\kappa^{\prime}(H_0 s_\alpha + A_0 c_\alpha)\overline{\mu}_R\psi_L 
\nonumber  \\ &
+ g^{\prime}_z\overline{\mu}_L\gamma^{\mu}\psi_L Z_{\mu} +g^{\prime}_w\overline{\nu}\gamma^{\mu}\psi_L W^{+}_{\mu} + \text{H.c.} \bigg] 
\label{Eq:psi-int}
\end{align}
where 
\begin{equation}
g^{\prime}_z = -\frac{g'_w}{\sqrt{2}c_W},\quad g^{\prime}_w = \frac{\kappa g}{2}\frac{v}{M_{\psi}}. 
\label{Eq:gzp-gwp}
\end{equation} 
Note that the $\{H_0,A_0\}\mu\psi$ terms in Eq.~\eqref{Eq:psi-int} 
can contribute to $\Delta a_{\mu}$ via one-loop Feynman diagrams~\cite{Hiller:2019mou}.

From Eq.~\eqref{eq:lagrangian} we can read off ten undetermined parameters in this model:
\begin{equation}
g_{\chi},\quad s_\alpha,\quad M_{\chi},\quad \lambda_{HA},\quad \mu^{\prime}_A ,
\quad \lambda_A,\quad \kappa,\quad \kappa^{\prime},\quad y,\quad M_{\psi}. 
\end{equation}
We can further fix $\mu^{\prime}_A= 5.0\gev$, $\lambda_A= 1.0$ and $y= 0.01$ for this analysis because 
their changes are neither relevant nor significant for this work. 
Considering that the LHC Higgs boson measurements constrain $\sin^2\alpha < 0.12$ at $95\%$
C.L.~\cite{Aad:2015pla,Khachatryan:2016vau},
we assign the upper bound $\sin\alpha < 0.3$ in our parameter scan\footnote{
We will see in the next section that constraints
from Higgs boson invisible and exotic decays are much stronger than this limit in our model.}. 
As required by Eq.~\eqref{Eq:VLML_mix}, $\kappa$ cannot be large and
one has to introduce a larger $\kappa^{\prime}$ to explain the $\Delta a_{\mu}$ deviation. 
Furthermore, a very massive $\psi^{\pm}$ could suppress the contribution to $\Delta a_{\mu}$ 
when we require perturbativity of $\kappa^{\prime}$ with $\kappa^{\prime}\leq\sqrt{4\pi}$. 
Finally, for simplicity, we assume $M_{A_0}/2 < M_{\chi} < M_{\psi}$ such that
$A_0\rightarrow\chi\overline{\chi}$ and the annihilation
$\chi\overline{\chi}\rightarrow \psi^{+}\psi^{-}$ are kinematically forbidden. 
Taking all these facts into consideration,
the scan range for the non-fixed parameters is given by the followings bounds
\begin{eqnarray}
10^{-3} &\leq \, g^{(*)}_{\chi}  \, \,\leq& 2.0 \; , \nonumber\\
5\times 10^{-3} &\leq \, \sin\alpha^{(*)}  \, \,\leq& 0.3 \; , \nonumber\\
150.0 &\leq \, M_{\psi} / \textrm{GeV} \, \leq & 450.0 \; , \nonumber\\
\label{domain} 
26.0 &\leq \, M_{\chi} / \textrm{GeV} \, \leq & 0.9\times M_{\psi}/\textrm{GeV} \; ,\\
5\times 10^{-4} &\leq \, \lambda_{HA}^{(*)} \, \,\leq& 10^{-2} \; , \nonumber\\ 
3\times 10^{-6} &\leq \, \kappa^{(*)} \, \,\leq& 8\times 10^{-2} \; , \nonumber\\
1.0 &\leq \, \kappa^{\prime} \,\leq& \sqrt{4\pi} \; , \nonumber
\end{eqnarray}
where the star $(*)$ indicates that the parameter is scanned logarithmically in base 10.

\section{Experimental constraints}
\label{sec:constraint}

Recently, the Muon $g-2$ Collaboration at Fermilab reported a new measurement
of the anomalous magnetic dipole moment of the muon achieving
the new combined result~\cite{Abi:2021gix}
\begin{equation}
\Delta a_{\mu}=(2.51\pm 0.59)\times 10^{-9}
\label{eq:E989}
\end{equation} 
consistent with a $4.2\sigma$ deviation from the SM\@.
Note that the SM uncertainties have been taken into account in
the error bar%
\footnote{The SM central value and error bar for $(g-2)_\mu$ use the 
    theoretical consensus value with hadronic vacuum polarization (HVP)
    determined from $e^+e^-\to \text{hadrons}$ (see, for example,~\cite{Davier:2019can}).
    Note that a recent calculation for the HVP on the lattice finds no significant
    tension between measurement and SM prediction~\cite{Borsanyi:2020mff}.
    }.
The dominant BSM contributions to $(g-2)_{\mu}$ in this model comes from one-loop Feynman diagrams
with $\psi^{\pm}$, $A_0$, and $H_0$ as~\cite{Hiller:2019mou} 
\begin{equation}
\Delta a_{\mu} = \frac{\kappa^{\prime 2}}{96\pi^2}\frac{m^2_{\mu}}{M^2_{\psi}}\left[ c_\alpha^2 f\!\left(\frac{M^2_{A_0}}{M^2_{\psi}}\right)+s_\alpha^2 f\!\left(\frac{M^2_{H_0}}{M^2_{\psi}}\right) \right] 
\label{Eq:mug2}
\end{equation}
where $f(t)=(2t^3 +3t^2 -6t^2 \text{ln}t -6t+1)/(t-1)^4$. 
Because of the constraint in Eq.~\eqref{Eq:VLML_mix} and large CP-violation effect of muon EDM, 
we can safely assume $\kappa\ll\kappa^{\prime}$.

Furthermore, the ATLAS collaboration has detected a local excess in 
the Higgs decay channel $H_0\to A_0 A_0\to b\overline{b}\mu^{+}\mu^{-}$ at $M_{A_0}=52$ GeV, 
by using 139~fb$^{-1}$ data at $\sqrt{s}=13$ TeV~\cite{ATLAS:2021ypo}. 
The measured branching ratio ${\rm BR}(H_0\rightarrow A_0 A_0\rightarrow b\overline{b}\mu^{+}\mu^{-})$ 
is around $3.5\times 10^{-4}$ with a local $3.3\sigma$ significance.
By taking a conservative approach, 
we assume that the central value is located at $3.5\times 10^{-4}$ 
and the likelihood is 
a Gaussian distribution with error bar equal to $3.5\times 10^{-4}$ divided by $3.3$
at $M_{A_0}=52$ GeV. 
In the on-shell limit, we can calculate 
${\rm BR}(H_0\rightarrow A_0 A_0)$, ${\rm BR}(A_0\rightarrow b\overline{b})$ and  
${\rm BR}(A_0\rightarrow\mu^{+}\mu^{-})$ individually and then multiply them together 
to obtain ${\rm BR}(H_0\rightarrow A_0 A_0\rightarrow b\overline{b}\mu^{+}\mu^{-})$.

We can divide the major constraints used for this study in five categories: 
(1) the LHC Higgs boson measurements, 
(2) the LEP and LHC $A_0$ searches,
(3) the DM phenomenology, 
(4) the ATLAS multi-lepton search and
(5) electron, muon electric dipole moments (EDMs).

\subsection{The LHC Higgs boson measurements}

For the LHC Higgs boson measurements, one can further classify them as 
\begin{itemize} 
\item \underline{\textbf{Higgs boson exotic and invisible decays}}\\
We take ${\rm BR}(H_0\rightarrow\text{undetected}) < 19\%$ and 
${\rm BR}(H_0\rightarrow\text{invisible}) < 9\%$ at $95\%$ C.L. as reported by Ref.~\cite{ATLAS:2020qdt}. 
The partial decay width of $H_0\rightarrow A_0 A_0$ can be written as 
\begin{equation}
\Gamma (H_0\rightarrow A_0 A_0) = \frac{\lambda^2_{H_0 A_0 A_0}}{32\pi M_{H_0}}\sqrt{1-4\left(\frac{M_{A_0}}{M_{H_0}}\right)^2} 
\label{Eq:H-exo}
\end{equation} 
where 
\begin{equation}
\lambda_{H_0 A_0 A_0} = -\mu_A s_\alpha^3 -2(3\lambda_H -2\lambda_{HA})v s_\alpha^2 c_\alpha 
-2\lambda_{HA}v c_\alpha^3 +(2\mu_A -\mu^{\prime}_A) 
s_\alpha c_\alpha^2.
\end{equation} 
If $m_{\chi} < M_{H_0}/2$, the Higgs boson can also decay to a pair of DM as 
\begin{equation}
\Gamma (H_0\rightarrow\chi\overline{\chi}) = s_\alpha^2 g^2_{\chi}\frac{M_{H_0}}{8\pi}\left(1-\frac{4m^2_{\chi}}{M^2_{H_0}}\right)^{3/2}. 
\label{Eq:H-DM}
\end{equation} 
Therefore, we require 
\begin{eqnarray}
	&& {\rm BR}(H_0\to A_0 A_0)+{\rm BR}(H_0\to \chi\overline{\chi})< 19\%,~{\rm and}\nonumber\\
	&& {\rm BR}(H_0\to \chi\overline{\chi})<9\%.
\end{eqnarray}
Note that, in this study, we fix $M_{A_0}=52$~GeV
such that $\Gamma (H_0\to A_0 A_0)$ is a function
of $\sin\alpha$ and $\lambda_{HA}$ only.

\item \underline{\textbf{$H_0\rightarrow\mu^{+}\mu^{-}$}}\\
The decay width $\Gamma (H_0\rightarrow\mu^{+}\mu^{-})$ can be modified 
by changing the $\mu$-$\psi$ mixing.
We require the measured signal strength of $H_0\rightarrow\mu^{+}\mu^{-}$,
relative to the SM prediction, to be $1.19^{+0.44}_{-0.42}(\text{stat})^{+0.15}_{-0.14}(\text{syst})$
as reported by the CMS collaboration~\cite{Sirunyan:2020two}.
In this work, we focus on the result from CMS because of the rather
larger error bar in the one from the ATLAS collaboration~\cite{Aad:2020xfq}.

\item \underline{\textbf{$H_0\rightarrow\gamma\gamma$}}\\
The VLML $\psi^{\pm}$ can also contribute $H_0\rightarrow\gamma\gamma$ and 
further reduce the SM prediction.
The partial decay width of $H_0\rightarrow\gamma\gamma$ in our model is 
\begin{equation}
\Gamma (H_0\rightarrow\gamma\gamma) = \frac{\alpha^2_{em}M^3_{H_0}}{256\pi^3 v^2} \left|c_\alpha \left(I^f_{H_0}(\tau_f)+ I^W_{H_0}(\tau_W) \right)+ s_\alpha I^{\psi}_{H_0}(\tau_{\psi})\right|^2
\end{equation} 
where 
\begin{align} 
& I^f_{H_0}(\tau_f) = -2 N_{cf} e^2_f \tau_f \left[1 +(1-\tau_f)f(\tau_f)\right]
\nonumber  \\ & 
I^W_{H_0}(\tau_W) = 2 +3\tau_W +3\tau_W (2-\tau_W) f(\tau_W) 
\nonumber  \\ & 
I^{\psi}_{H_0}(\tau_{\psi}) = - e^2_{\psi}\frac{4y}{g}\frac{m_W}{M_{\psi}}\tau_{\psi}f(\tau_{\psi})
\end{align}
with $\tau_i = 4m^2_i /M^2_{H_0}$ and 
\begin{eqnarray}
f(\tau) = \left\{
       \begin{array}{lr}
   \left(\sin^{-1}\sqrt {1/\tau} \right)^2 , & \quad \tau \ge 1
   \\
    - \frac{1}{4}\left[\ln \left(\frac{{1 + \sqrt {1 - \tau} }}
      {{1 - \sqrt {1 - \tau} }}\right) - i \pi \right]^2 , & \quad \tau < 1
\end{array}
      \right.
\end{eqnarray} 
where $\alpha_{em}$ is the fine-structure constant and $N_c = 3 (1)$ for quarks (charged leptons).
We require the measured signal strength of $H_0\rightarrow\gamma\gamma$, relative to the SM prediction, to be $1.12^{+0.07}_{-0.06}(\text{stat})^{+0.06}_{-0.07}(\text{syst})$ from CMS collaboration~\cite{Sirunyan:2021ybb}.
\end{itemize}

\subsection{The LEP and LHC $A_0$ searches}

The pseudoscalar $A_0$ with a mass of 52~GeV can be explored at the LEP and LHC experiments. 
The decay modes of $A_0$ for these searches are $A_0\rightarrow b\overline{b}, \tau^{+}\tau^{-}, \mu^{+}\mu^{-}$. 
For the search at LEP, the most stringent limit is 
$\sin^2\alpha\times \text{BR}(A_0\rightarrow b\overline{b})\lesssim 3.5\times 10^{-2}$ 
from the $e^{+}e^{-}\rightarrow Z A_0$ production~\cite{Barate:2003sz}. 
On the other hand, the multilepton final states from the LHC search 
$p p\rightarrow t \overline{t}\left( A_0\rightarrow\mu^{+}\mu^{-}\right)$ 
can provide a strong constraint on 
$\sin^2\alpha\times \text{BR}(A_0\rightarrow \mu^{+}\mu^{-})\lesssim 5\times 10^{-3}$~\cite{CMS:2019xud}. 
Finally, our model is unconstrained in the $A_0\rightarrow\tau^{+}\tau^{-}$ channel 
where $A_0$ is produced via $p p\rightarrow b\overline{b}A_0$~\cite{CMS:2019hvr}. 
We will see in Sec.~\ref{sec:result} that
the allowed range for $\sin\alpha$ remains smaller than these $A_0$ limits from LEP and LHC. 

\subsection{The DM phenomenology}

The DM phenomenology can be classified in three parts:
\begin{itemize}
\item \underline{\textbf{DM relic density}}\\
Since we fixed the mass of $A_0$ to be $M_{A_0}=52\gev$ and the annihilation channel 
$\chi\overline{\chi}\to \psi^{+}\psi^{-}$ is kinematically forbidden, 
one would expect that the Higgs resonance regions $\chi\overline{\chi}\to f\overline{f}$ can play an important role 
to enhance the annihilation cross section in the early universe. 
However, the Higgs exchange is suppressed by $\sin\alpha$ when compared with 
$A_0$ exchange and 
the dominant channel contributing to the relic density is $\chi\overline{\chi}\to \mu^\pm\psi^\mp$ 
via $A_0$ exchange when it is open. 
The most favorable size of $g_\chi \kappa^\prime$ in Eq.~\eqref{Eq:psi-int} 
can be determined using Planck's relic density measurement 
$\Omega_\chi h^2=0.12\pm 0.001$~\cite{Aghanim:2018eyx}. 
For example, a value of $\kappa^\prime\simeq 2.46$ with $g_\chi\simeq 0.13$ 
can fulfill the Planck constraint for $m_\chi$ around $200$~GeV. 

\item \underline{\textbf{DM direct detection}}\\
In Eq.~\eqref{Eq:HA-int}, DM interacts with quarks via $H_0/A_0$ exchange resulting in
a suppressed tree-level amplitude for DM-nucleon elastic scattering
due to small momentum transfer. 
With a simple estimation based on Refs.~\cite{Arcadi:2017wqi,Abe:2018emu}, 
we find that the loop contribution with the condition $g_\chi\sin\alpha\lesssim 0.05$
is still below the neutrino floor.
Our upper limit $g_\chi\sin\alpha\lesssim 0.6$
could bring the loop contribution above the neutrino floor
but is still well below the current XENON1T limit~\cite{Aprile:2018dbl}. 
Note that the complete two-loop DM-nucleon elastic scattering
can be found in Ref.~\cite{Ertas:2019dew}.
Our model corresponds to the special case
with CP phases $\phi_{\chi}=\pi/2$ and $\phi_\text{SM}=0$
in Eq.~(2.1) of the aforementioned reference.
They have shown that the full two-loop calculation
can lead to a smaller cross section
than previous approaches~\cite{Arcadi:2017wqi,Abe:2018emu}.
As we will see in Sec.~\ref{sec:result}, our actual result may be below $g_\chi\sin\alpha = 0.05$.
Hence, we simply ignore DM direct detection constraints in this study. 
\item \underline{\textbf{DM indirect detection}}\\ 
The dominant channel of DM annihilation in the present universe within our explored parameter space 
is still $\chi\overline{\chi}\to \mu^\pm\psi^\mp$ which is $s$-wave annihilation. 
The VLML $\psi^\mp$ can successively decay to $\mu^\mp b\overline{b}$. 
The final state $\mu^+\mu^- b\overline{b}$ can produce a soft photon or electron spectrum. 
Therefore, one may indirectly detect DM annihilation by using dSphs gamma ray data from 
Fermi~\cite{Ackermann:2015zua} and electron-positron data from AMS02~\cite{Aguilar:2021tos}.  
In addition, the produced photons and relativistic electrons may ionize and heat 
the universe gas at the recombination epoch.  
Thus, one may constrain DM annihilation cross section 
$\sv$ by using Cosmic Microwave Background (CMB) anisotropy data from Planck~\cite{Aghanim:2018eyx}. 
In this study, because the cosmic ray propagation is rather uncertain 
and CMB limit is weaker than the one from Fermi dSphs gamma ray data---in our DM mass range
of interest---%
we will only include the Fermi dSphs data in our analysis.
\end{itemize}

\subsection{The ATLAS multi-lepton search}

\begin{figure}[htbp]
\begin{centering}
\includegraphics[width=0.47\textwidth]{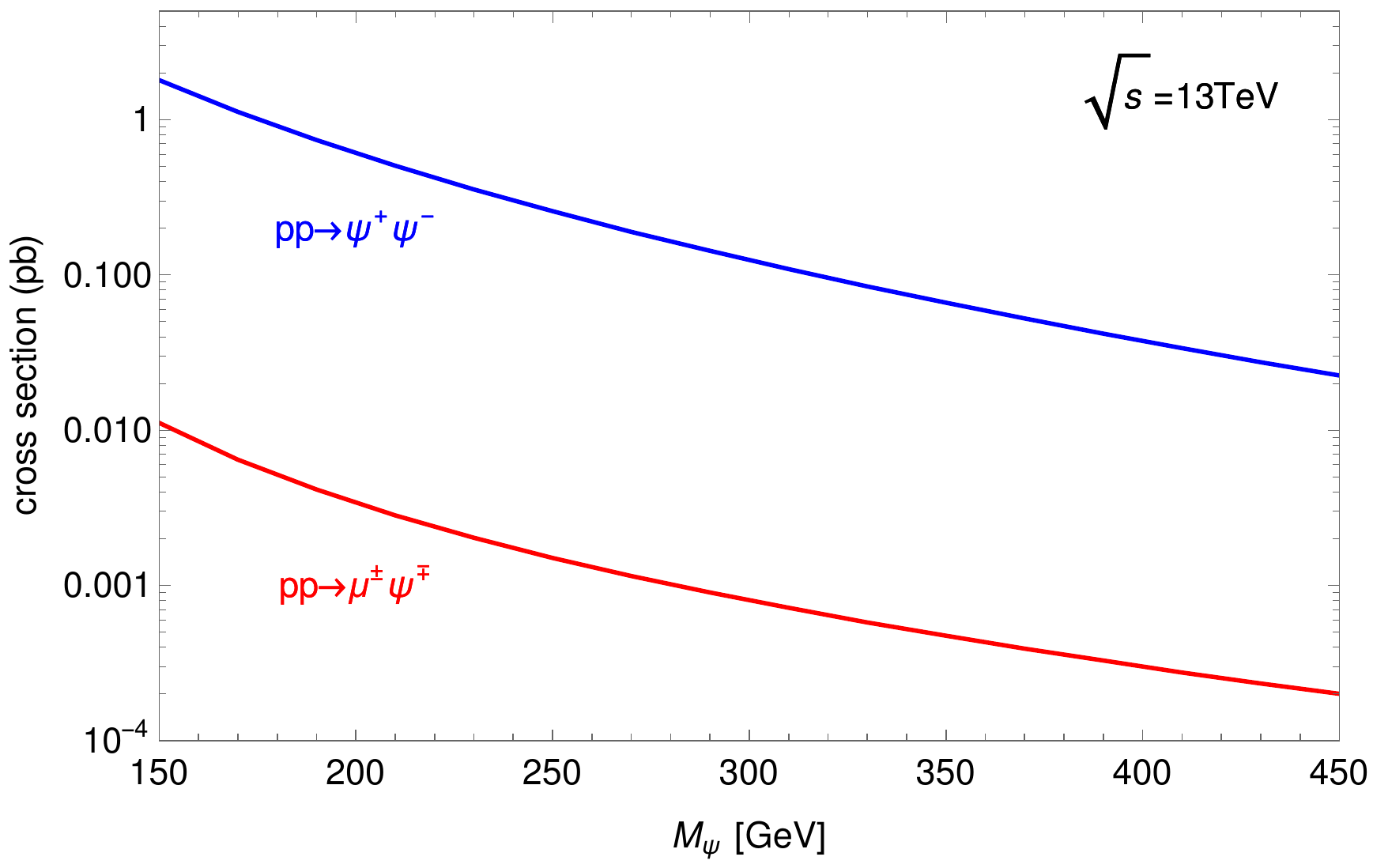} 
\includegraphics[width=0.47\textwidth]{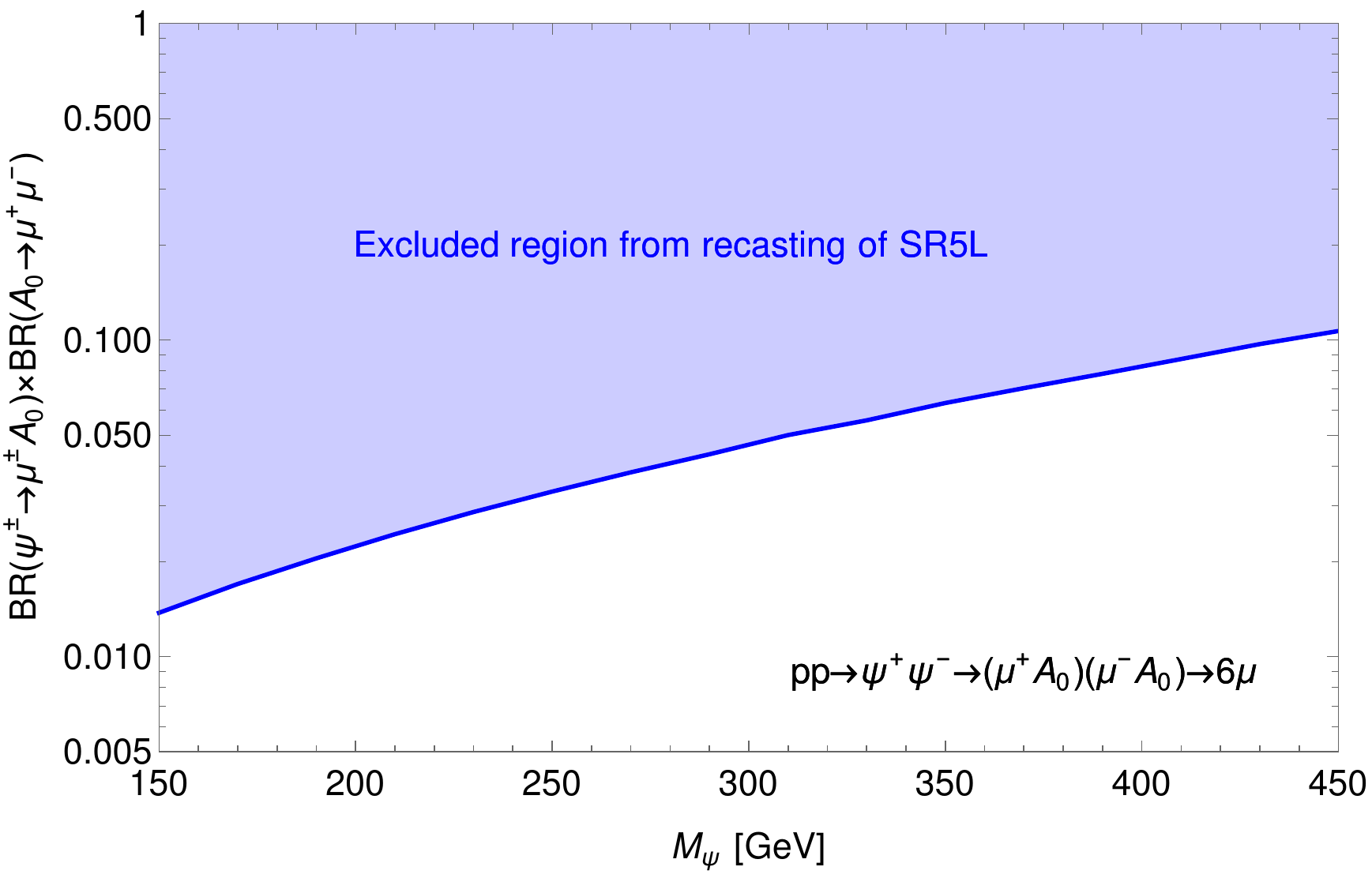}
\caption{ 
Left panel: The production cross sections for the VLML $\psi^{\pm}$ at $\sqrt{s}=13$ TeV. We fix $\sin\alpha = 0.1$, $\kappa = 5\times 10^{-2}$ and $\kappa^{\prime} = 2.0$ but vary $M_{\psi}$ from $150-450$ GeV. 
Right panel: Exclusion limit from recasting of the signal region \textbf{SR5L} in~\cite{Aad:2021qct} on $(M_{\psi},BR(\psi^{\pm}\rightarrow\mu^{\pm}A_0)\times BR(A_0\rightarrow\mu^{+}\mu^{-}))$ plane.
}
\label{fig:VLML_Xsec}
\end{centering}
\end{figure}

The VLML $\psi^{\pm}$ with $150\leq M_{\psi}\leq 450$~GeV
in our model is in a mass range that can be explored at the LHC via single and pair productions~\cite{Bissmann:2020lge}. 
The major single production of $\psi^{\pm}$ is via $g g\rightarrow H^{\ast}_0/A^{\ast}_0\rightarrow\mu^{\pm}\psi^{\mp}$ process. 
Since the allowed value of $\kappa$ is tiny, the single production from $pp\rightarrow W^{\pm\ast}\rightarrow\nu\psi^{\pm}$ and $pp\rightarrow Z^{\ast}\rightarrow\mu^{\pm}\psi^{\mp}$ are highly suppressed regarding Eq.~(\ref{Eq:gzp-gwp}). 
Alternatively, the major pair production of $\psi^{\pm}$ is via $pp\rightarrow \gamma^{\ast}, Z^{\ast}\rightarrow\psi^{+}\psi^{-}$ process.
We have checked that the pair production from
$gg\rightarrow H^{\ast}_0/A^{\ast}_0\rightarrow\psi^{+}\psi^{-}$
is much smaller than the Drell-Yan type process with our parameter settings. 
The production cross sections of Drell-Yan type process are only dependent on $M_{\psi}$, but the single production process is a function of $M_{\psi}$, $\sin\alpha$, $\kappa$ and $\kappa^{\prime}$. Therefore, 
we fix $\sin\alpha = 0.1$, $\kappa = 5\times 10^{-2}$ and $\kappa^{\prime} = 2.0$
but vary $M_{\psi}$ from $150$--$450$~GeV to show both single and pair productions at $\sqrt{s}=13$~TeV in the left panel of Fig.~\ref{fig:VLML_Xsec}. 
Since the cross sections of pair production are much larger than single production ones, we will focus on the constraint of ATLAS multi-lepton search~\cite{Aad:2021qct} on the pair production channel in this analysis. 
We remark that $\psi^{\pm}\rightarrow\mu^{\pm}A_0$ is the dominant decay mode (${\rm BR}(\psi^{\pm}\rightarrow\mu^{\pm}A_0)\gtrsim 99\%$)
for our parameter settings. 

We closely follow the simulation and analysis in ATLAS multi-lepton search~\cite{Aad:2021qct} 
based on a data sample with ${\cal L}=139~{\rm fb}^{-1}$ at $\sqrt{s}=13$~TeV. 
The package \textbf{MadAnalysis 5}~\cite{Conte:2014zja,Dumont:2014tja} is used to recast the signal region \textbf{SR5L} for the following process 
\begin{equation}
pp\rightarrow \gamma^{\ast}, Z^{\ast}\rightarrow\psi^{+}\psi^{-}\rightarrow (\mu^{+}A_0)(\mu^{-}A_0)\rightarrow 6\mu. 
\label{Eq:6mu}
\end{equation} 
First, the signal processes in Eq.~(\ref{Eq:6mu}) with up to two extra partons were generated from the 
leading order matrix elements by using
\textbf{Madgraph5 aMC@NLO}~\cite{Alwall:2014hca}.
We apply Mangano's prescription (MLM)~\cite{Mangano:2006rw,Alwall:2007fs} for jet-parton matching with $M_{\psi}/4$ as the matching scale\footnote{ 
The $K$-factor (QCD corrections) for Drell-Yan type process ranges between $1.2$ and $1.5$. Since its effect is mild, we will not include this $K$-factor in our recasting.}. 
The \textbf{Pythia8}~\cite{Sjostrand:2007gs} is used for parton showering and merging 
as well as hadronization. 
We have modified the ATLAS template in \textbf{Delphes3}~\cite{deFavereau:2013fsa} to be consistent 
with the setup of Ref.~\cite{Aad:2021qct} for fast detector simulation. 
The event selections in this analysis are summarized below:  
\begin{itemize} 
\item First, in order to suppress background muons from semileptonic decays of $c$- and $b$-hadrons, any muon within $\Delta R = 0.4$ of a jet is removed. 
\item One of the following triggers with efficiency $\geq 90\%$ is used: (1) $P_T(\mu_1) > 27$~GeV, (2) $P_T(\mu_{1,2}) > 15$~GeV and (3) $P_T(\mu_1) > 23$~GeV, $P_T(\mu_2) > 9$~GeV where the subscript of muon is in the order of $P_T$.
\item The signal muons must have $N(\mu)\geq 5$ with $P_T(\mu) > 5$~GeV and $|\eta(\mu)| < 2.7$.
\item $b$-jet veto 
\item hadronic $\tau$ veto 
\item Furthermore, to suppress low-mass particle decays from backgrounds, $m_{\mu^{+}\mu^{-}} > 4$~GeV and $8.4 < m_{\mu^{+}\mu^{-}} < 10.4$~GeV veto are required. 
\item Finally, if two muons are found in $\Delta R = 0.6$ and $P_T(\mu) < 30$~GeV for one of them, both muons are discarded. This selection is used to suppress leptons from a decay chain with multiple heavy flavour quarks backgrounds.   
\end{itemize} 
The model-independent limit
is reported as $\langle\epsilon\sigma\rangle^{95}_{obs} = 0.129$~fb
as calculated at $95\%$~C.L. from the signal region \textbf{SR5L}\footnote{
As pointed out in~\cite{Aad:2021qct}, there is an excess over the SM background
in the signal region \textbf{SR5L} with the local significance $1.9\sigma$.
Similar to~\cite{Kawamura:2021ygg}, our model can also explain this mild excess if it is confirmed in the future.}. 
We apply this limit to constrain BR$(\psi^{\pm}\rightarrow\mu^{\pm}A_0)\times \textrm{BR}(A_0\rightarrow\mu^{+}\mu^{-})$ for varying $M_{\psi}$ as shown in the right panel of Fig.~\ref{fig:VLML_Xsec}. It is expected to find that BR$(A_0\rightarrow\mu^{+}\mu^{-})$ already suffers from strong constraints such that $A_0$ cannot be muonphilic.

\subsection{The EDM of electron and muon}

On the EDM side, the latest electron EDM measurement from the ACME collaboration
is reported as $|d^E_e| < 1.1\times 10^{-29}~e$cm 
at $90\%$ C.L.~\cite{Andreev:2018ayy}. In the case of the muon EDM, the measurement from the Muon $g-2$ Collaboration 
is {$|d^E_{\mu}| < 1.9\times 10^{-19}~e$cm} at $95\%$ C.L.~\cite{Bennett:2008dy}.  
The detailed formulas for both electron and muon EDMs in this model are given in Appendix~\ref{sec:append}. 
We adopt both electron and muon EDMs constraints in our analysis. Note that the constraints 
from tau and neutron EDMs from the Belle collaboration~\cite{Inami:2002ah} and Ref.~\cite{Afach:2015sja}, respectively,
are much weaker than electron and muon EDM measurements in this model
and can be safely ignored.

\section{Results}
\label{sec:result}

In this section we will present our numerical results. 
In Eq.~\eqref{domain}, we show our prior ranges for the model parameters 
while in Sec.~\ref{sec:constraint} all the observables that are used in our analysis are discussed.
The corresponding model file required for DM phenomenology
is generated with \textbf{FeynRules}~\cite{Alloul:2013bka}. 
The DM relic density, annihilation cross section, and Higgs decay width at tree-level 
are computed by using \textbf{micrOMEGAs}~\cite{Belanger:2020gnr}. 
We perform our global scan with the Markov Chain Monte Carlo package \textbf{emcee}~\cite{ForemanMackey:2012ig}. 

To determine the allowed parameter space at 1$\sigma$ and 2$\sigma$ we analyse
the data obtained via \textbf{emcee} using the profile likelihood (PL)
method.
This approach allows us to remove the unwanted parameters as nuisance
parameters by maximizing the likelihood over them.
After profiling  the unwanted parameters we are left with the likelihood of
the parameters we are interested in.
The likelihood for $n$ parameters of interest, $\mathcal{L}(x_1, x_2, \ldots,x_n)$, can be integrated as
\begin{equation}
\frac{1}{\mathcal{N}}\int_\mathcal{C} \mathcal{L}(x_1, x_2, \ldots, x_n)dx_1 dx_2\ldots dx_n = \varrho
\end{equation}
where $\mathcal{C}$ is the smallest $n$-volume with probability $\varrho$,
$x_k$ are placeholder parameters and $1/\mathcal{N}$ is a normalization factor
with $\mathcal{N}$ the result of integrating $\mathcal{L}$ inside $\mathcal{C}\to\infty$.

Our likelihood function can be modeled as a pure Gaussian distribution
using the total $\chi_\text{total}^2$ in the form
\begin{equation}
\mathcal{L} = \exp(-\Delta\chi_\text{total}^2/2),\qquad
\Delta\chi_\text{total}^2 = \chi_\text{total}^2 - \min(\chi_\text{total}^2).
\end{equation}
In this study $\chi_\text{total}^2$ is made with the combination of the constraints mentioned
in Sec.~\ref{sec:constraint} and the Appendix~\ref{sec:append}:
\begin{align}
\chi_\text{total}^2 = {}& \chi^2(H_0\to A_0 A_0\to b\overline{b}\mu^+\mu^-) + \chi^2(\Delta a_\mu) + \chi^2(\Omega h^2)\nonumber\\
    &+ \chi^2(H_0\text{ decays}) + \chi^2(\mu,e\text{EDM}).
\end{align}

In the following subsections, we will discuss our result based on the $1\sigma$ and $2\sigma$ 
allowed regions of the previous BNL $(g-2)_{\mu}$ data (orange unfilled contours) and 
the combined results from BNL and the recent E989 (blue filled contours). 
However, all other constraints given in Sec.~\ref{sec:constraint} and Appendix~\ref{sec:append} are applied. 

\subsection{The impact from $(g-2)_{\mu}$ results on $\kappa^{\prime}$ and $M_{\psi}$}
\label{sec:result_g2}

\begin{figure}[htbp]
\begin{centering}
\includegraphics[width=0.6\textwidth]{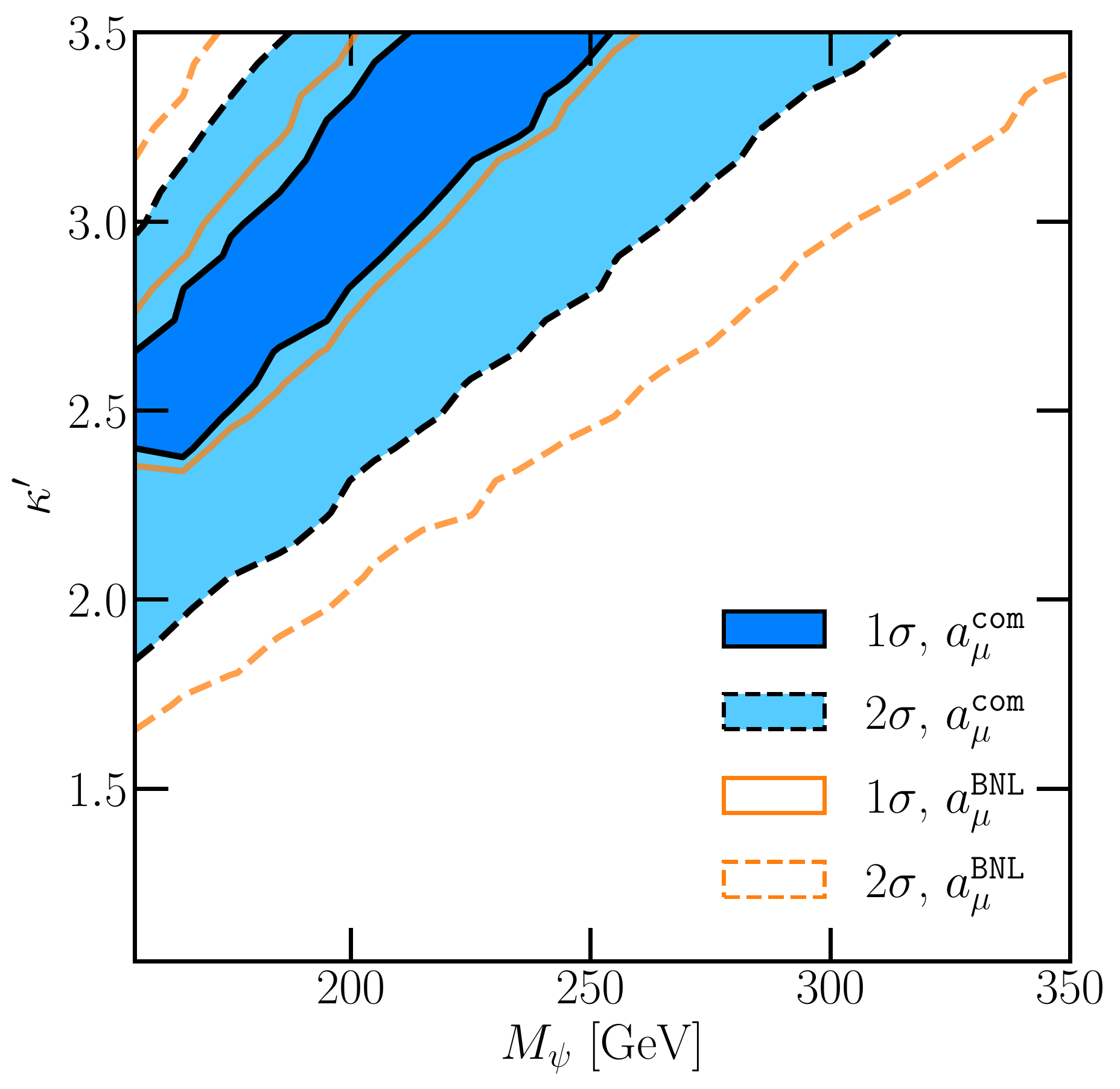} 
\caption{
The VLML mass $M_{\psi}$ {\it {vs.}} $\kappa^{\prime}$. 
The inner and outer contours are $1\sigma$ (solid) and $2\sigma$ (dashed) confidence limits, respectively. 
The orange unfilled and blue filled contours correspond to the previous BNL $(g-2)_{\mu}$ data 
and the combined one from BNL and the recent E989, respectively. 
In addition to $(g-2)_{\mu}$, all the other constraints are considered. 
}
\label{fig:mpsi_kapp}
\end{centering}
\end{figure}

From Eq.~\eqref{Eq:mug2},
after fixing $M_{H_0}$ and $M_{A_0}$,
it is easy to note that the dependence of $\Delta a_\mu$ is reduced to the parameters $\kappa^{\prime}$, $M_{\psi}$, and 
$\sin\alpha$. 
Furthermore, when compared with $M_{\psi}$ and $\kappa^{\prime}$, 
smaller values of the parameter $\sin\alpha$ do not have an important effect on
the value of $\Delta a_{\mu}$
due to the dominant term with $\cos\alpha$. 
In Fig.~\ref{fig:mpsi_kapp}, we present the correlation between $M_{\psi}$ and $\kappa^{\prime}$. 
For heavier $\psi^\pm$, a larger $\kappa^{\prime}$ is required to satisfy 
the measured value of $\Delta a_{\mu}$. 
It is clear that, when the new combined $(g-2)_{\mu}$ data is applied, 
the $2\sigma$ shrinks significantly while the $1\sigma$ region changes slightly. 
The new combined $(g-2)_{\mu}$ data constrains these parameters to the limits  
$\kappa^{\prime}>1.8$ and $M_{\psi}<315\gev$ with an almost linearly correlation
between them. 
In Sec~\ref{sec:LHC} we will describe how these facts can help in searches for
the VLML $\psi^{\pm}$ at the LHC\@.

\subsection{The impact from Higgs measurements on $\sin\alpha$, $\lambda_{HA}$, and $\kappa$}

\begin{figure}[htbp]
\begin{centering}
\includegraphics[width=0.49\textwidth]{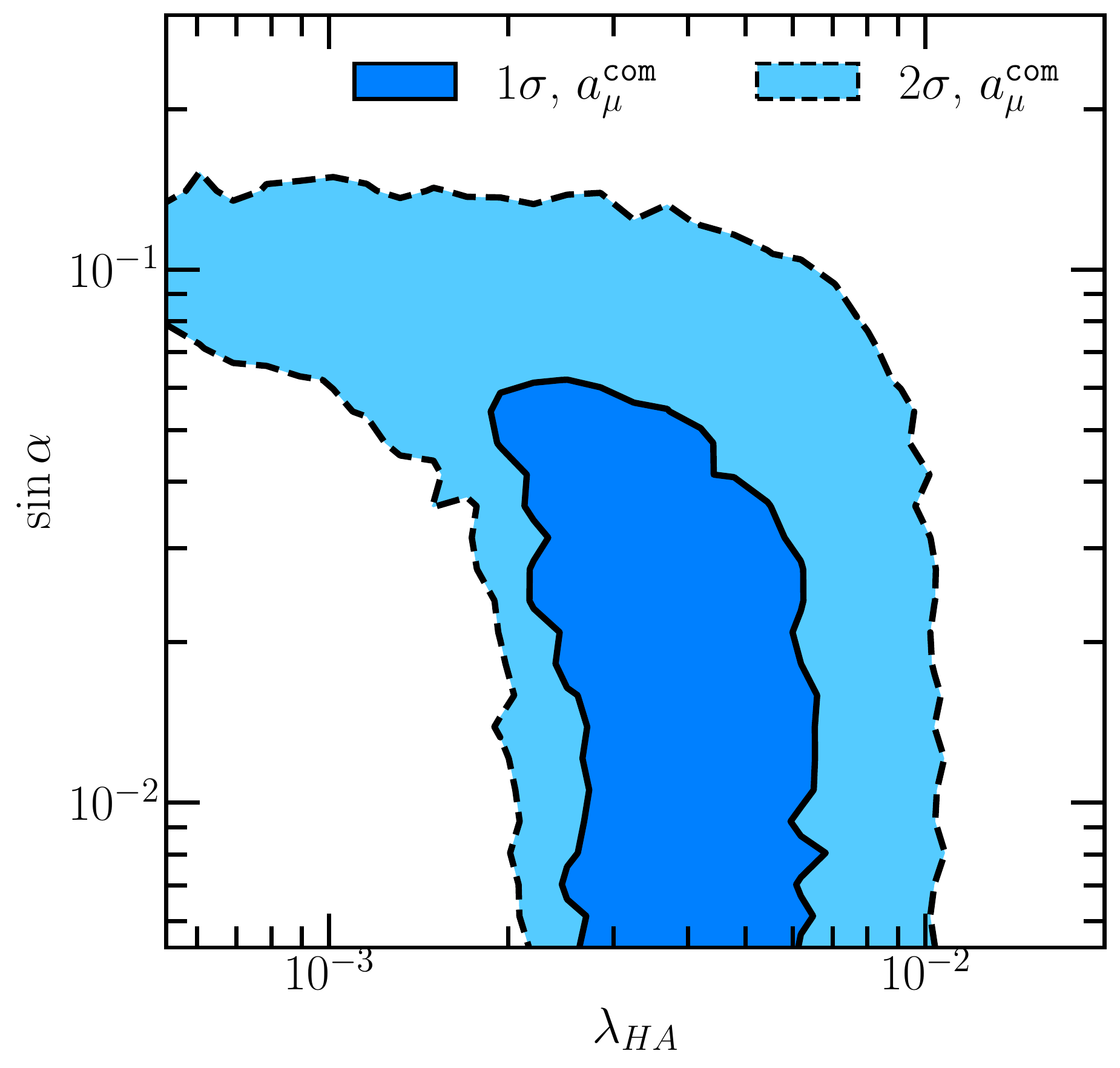} 
\includegraphics[width=0.49\textwidth]{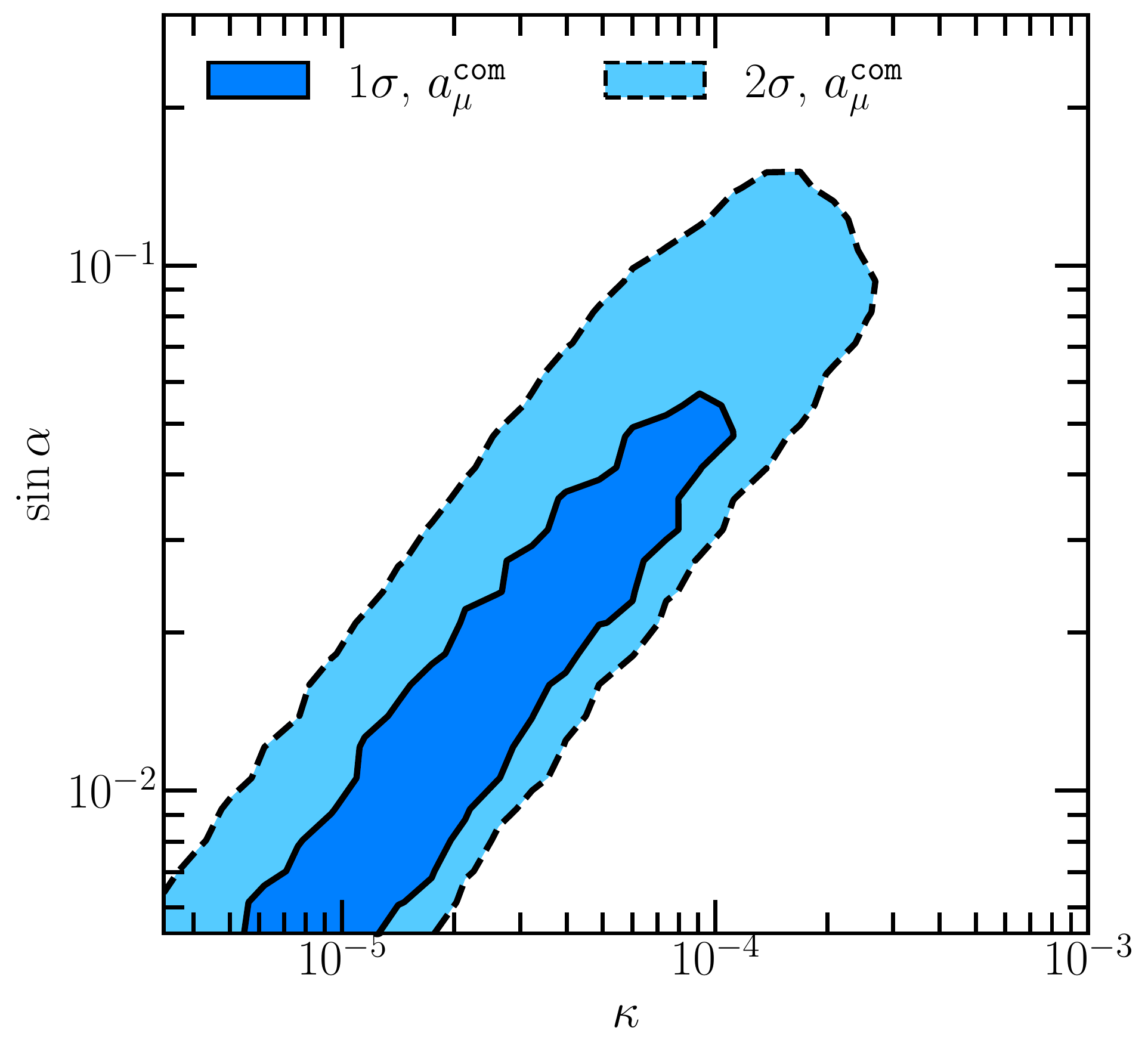} 
\caption{
The likelihood distribution in the ($\lambda_{HA}$, $\sin\alpha$) plane (left) 
and the ($\kappa$, $\sin\alpha$) plane (right). 
Colors and constraints are as described in Fig.~\ref{fig:mpsi_kapp}.
}
\label{fig:Higgs}
\end{centering}
\end{figure}

In Fig.~\ref{fig:Higgs}, we can clearly see that the difference between the blue and orange contours 
is small. As mentioned in the previous section, $\gtwo$ is not sensitive to $\sin\alpha$, $\lambda_{HA}$, 
and $\kappa$. On the other hand, the Higgs boson phenomenology is rather interesting for these three parameters. 
First of all, the Higgs boson exotic and invisible decay widths are functions of 
both $\sin\alpha$ and $\lambda_{HA}$ as can be seen in Eq.~\eqref{Eq:H-exo} and~\eqref{Eq:H-DM}.
From the left panel of Fig.~\ref{fig:Higgs} we can see the limits $\sin\alpha\lesssim 0.15$ and $\lambda_{HA}\lesssim 0.01$ . 
Trying to be consistent with the ATLAS data for Higgs decaying to $2b 2\mu$, 
we find that $\kappa$ is proportional to $\sin\alpha$ as presented in the right panel of Fig.~\ref{fig:Higgs}.

\begin{figure}[htbp]
\begin{centering}
\includegraphics[width=0.6\textwidth]{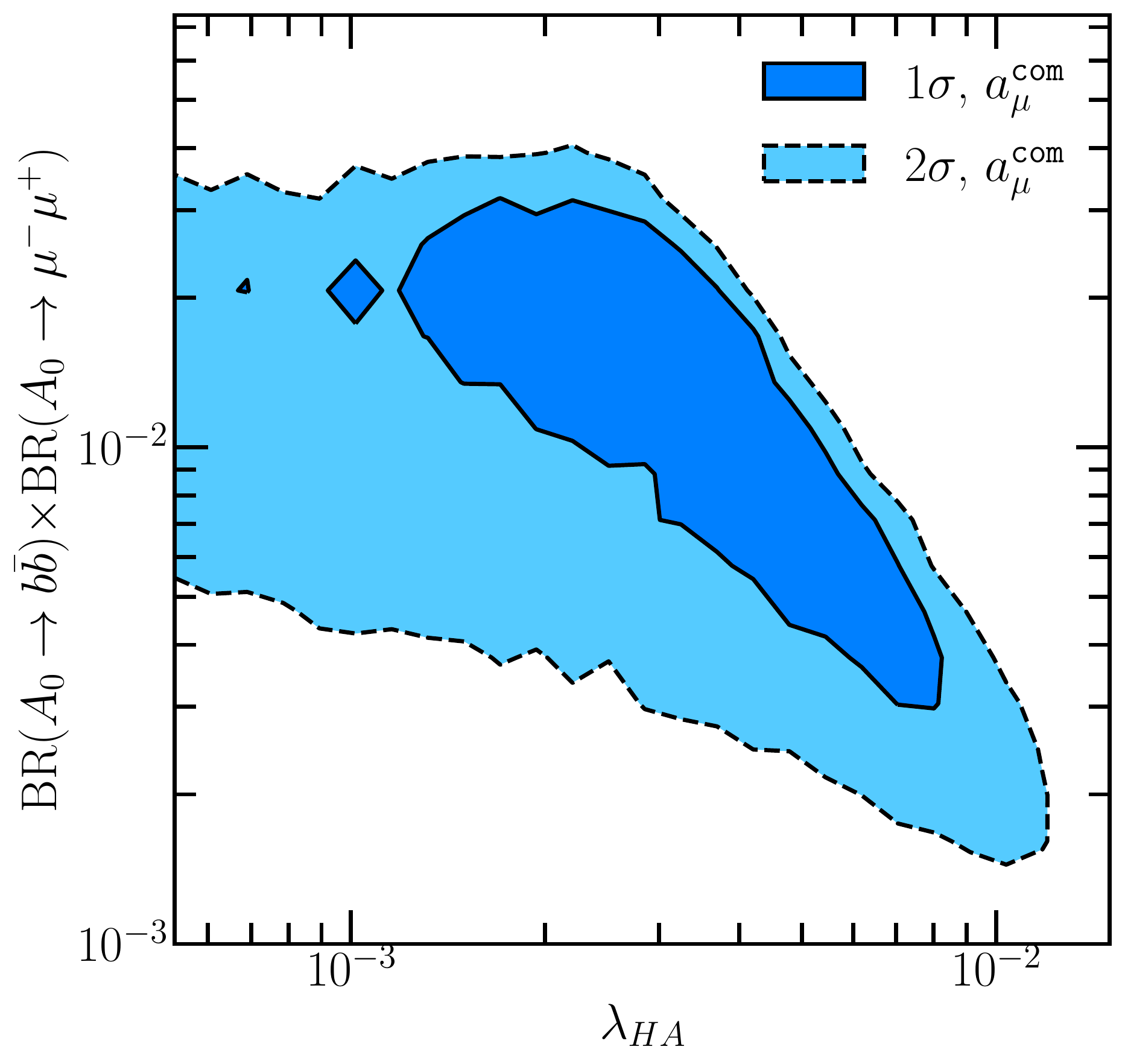} 
\caption{
The two dimensional likelihood contour in the plane ($\lambda_{HA}$, 
${\rm BR}(A_0\rightarrow b\overline{b})\times {\rm BR}(A_0\rightarrow\mu^{+}\mu^{-})$).
Colors and constraints are as described in Fig.~\ref{fig:mpsi_kapp}. 
}
\label{fig:BR}
\end{centering}
\end{figure}

Considering that the decay width of $A_0$ to any fermion except the muon changes
only by a constant mass factor and that all BRs add up to 1,
we expect an universal relationship between
BR($A_0 \to \mu^+\mu^-$) and BR($A_0\to f\overline{f}$) to be linear.
The reason is that
if ${\rm BR}(A_0\rightarrow b\overline{b})$ increases then
${\rm BR}(A_0\rightarrow\mu^{+}\mu^{-})$ decreases rapidly,
before BR($A_0\rightarrow b\overline{b}$) reaches 1.   
One can see, from Eq.~\eqref{Eq:gA}, that 
enhancing ${\rm BR}(A_0\rightarrow\mu^{+}\mu^{-})$ requires larger $\kappa$, 
while we also need to increase $\sin\alpha$ 
in order to satisfy ATLAS $2b2\mu$ measurement, which depends on 
${\rm BR}(A_0\rightarrow b\overline{b})\times {\rm BR}(A_0\rightarrow\mu^{+}\mu^{-})$.

In Fig.~\ref{fig:BR}, we show the $2\sigma$ (dashed) and $1\sigma$ (solid) 
favored regions on ($\lambda_{HA}$, 
${\rm BR}(A_0\rightarrow b\overline{b})\times {\rm BR}(A_0\rightarrow\mu^{+}\mu^{-})$) plane.  
When the product of both BRs is in the range $10^{-3}$--$4\times 10^{-2}$,
we find the limit
$0.85 < {\rm BR}(A_0\rightarrow b\overline{b}) < 0.89$.
Finally, $\kappa$ and $\sin\alpha$ are also constrained by other searches such as 
LHC Higgs decay ($H_0\rightarrow\mu^{+}\mu^{-}$ and $H_0\rightarrow\gamma\gamma$), 
and EDMs constraints (both electron and muon). Therefore, their values are restricted to be small
as can be seen in Fig.~\ref{fig:Higgs}.

\subsection{The impact from DM measurements on $g_{\chi}$, $M_{\chi}$, $M_{\psi}$, and $\sin\alpha$}

\begin{figure}[htbp]
\begin{centering}
\includegraphics[width=0.475\textwidth]{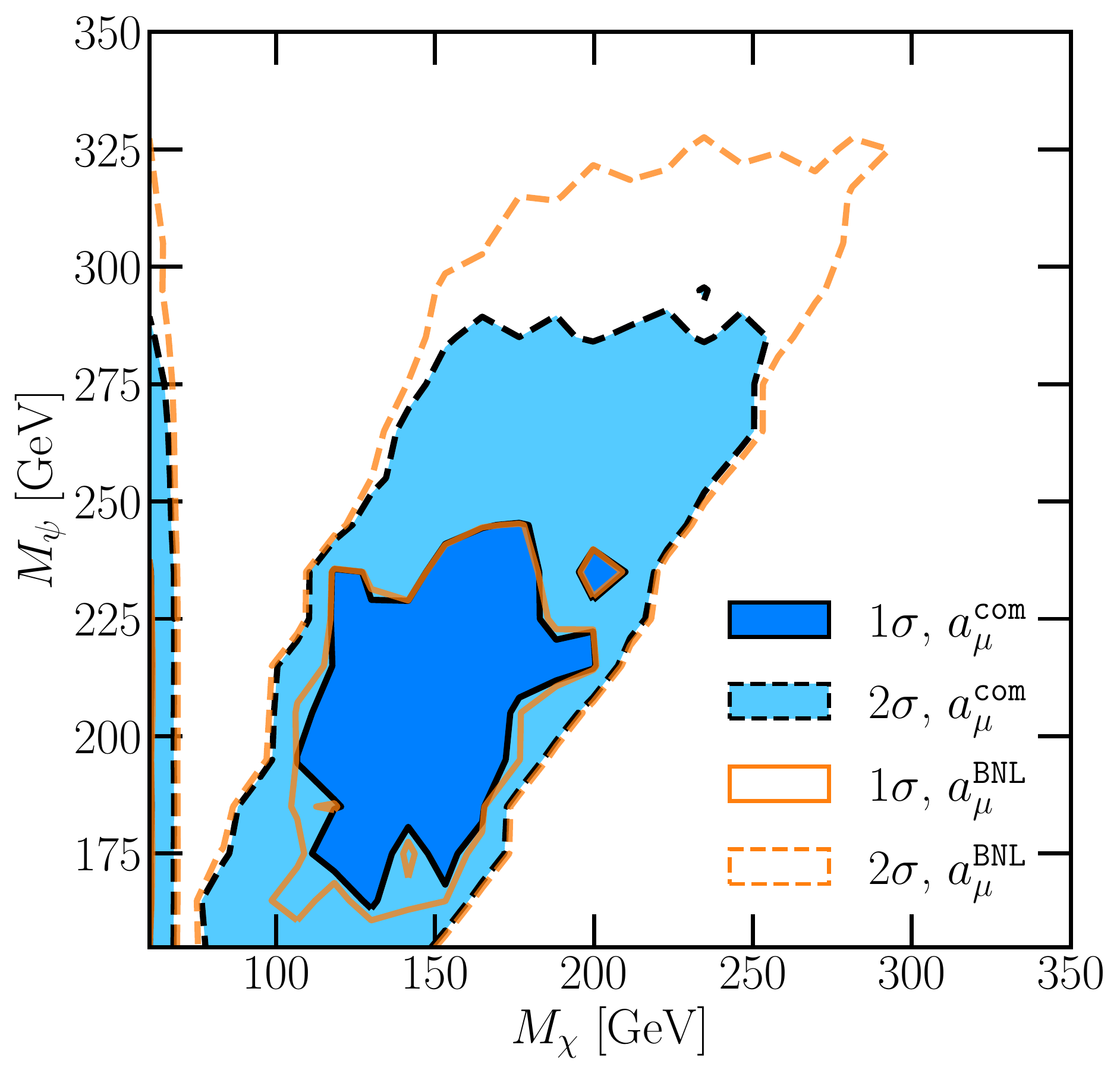}
\includegraphics[width=0.49\textwidth]{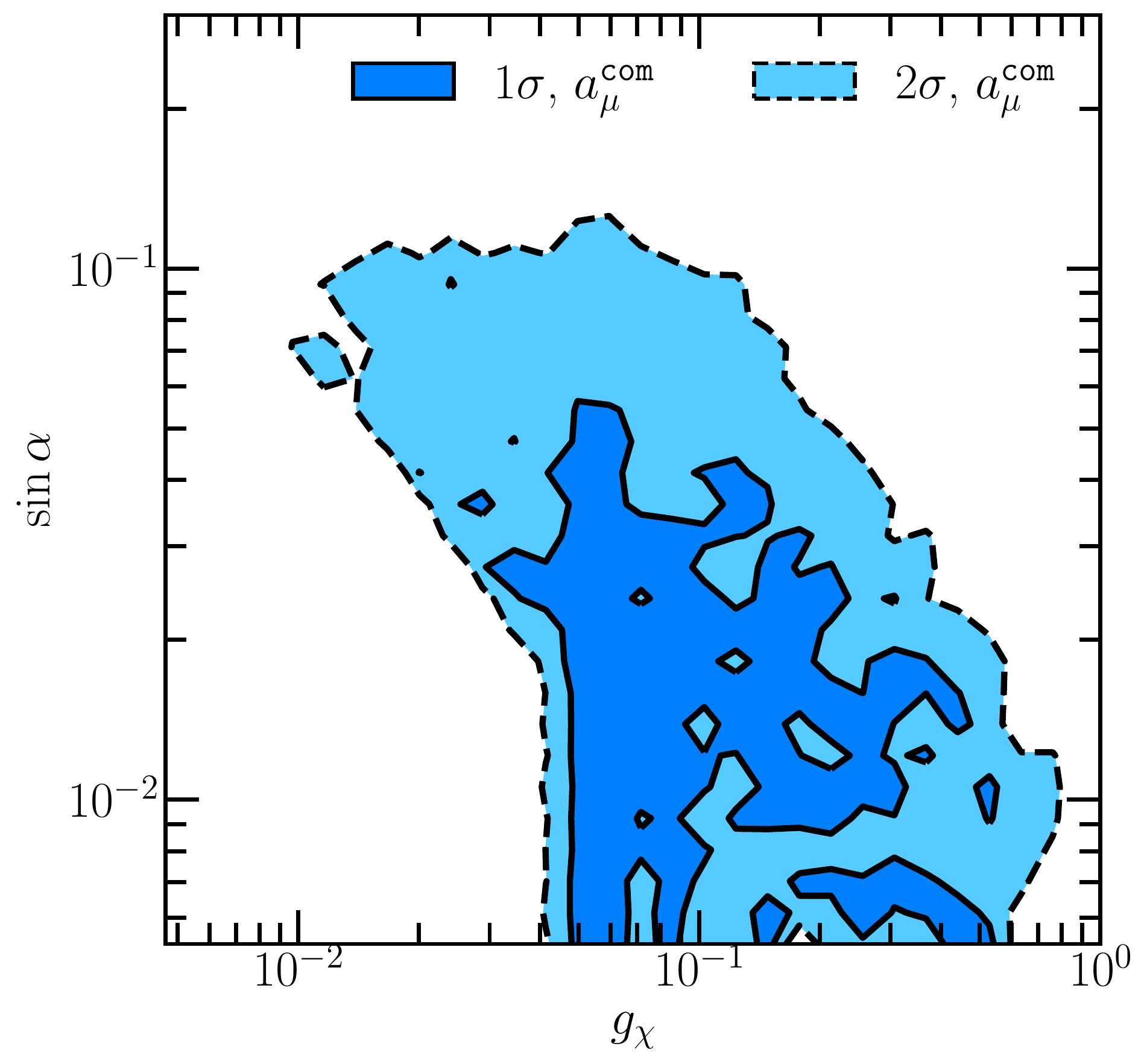}
\caption{
Likelihood distribution in the ($M_{\chi}, M_{\psi}$) plane (left) and
($g_{\chi},  \sin\alpha$) plane (right).
Colors and constraints are as described in Fig.~\ref{fig:mpsi_kapp}.
}
\label{fig:123}
\end{centering}
\end{figure}

In general, the relevant model parameters for DM phenomenology are 
$g_{\chi}$, $M_{\chi}$, $\kappa^{\prime}$, $M_{\psi}$ and $\sin\alpha$.
Since the allowed range of $\kappa^{\prime}$ is already restricted to
$1.8 < \kappa^{\prime}\lesssim\sqrt{4\pi}$,
here we focus on the analysis of $g_{\chi}$, $M_{\chi}$, $M_{\psi}$ and $\sin\alpha$.
In the left panel of Fig.~\ref{fig:123}, 
we can clearly observe that both ranges of $M_{\chi}$ and $M_{\psi}$ shrink if  
the new measurement of $(g-2)_{\mu}$ is applied. 
Note that only $\psi^{\pm}$ enters to the $(g-2)_{\mu}$ loop calculation,   
but DM mass $M_{\chi}$ and $M_{\psi}$ can be further restricted by Planck relic density constraint.  
In this region (larger $M_\chi$), the dominant channel of DM annihilation is $\chi\overline{\chi}\to \mu^\pm\psi^\mp$.

For the Higgs resonance region, where the $\chi\overline{\chi}\to f\overline{f}$ is relevant,
the presence of terms with $g_{\chi}\sin\alpha$ will bring some regions with large
$g_\chi\sim \mathcal{O}(10^{-1})$ into the 2$\sigma$ and 1$\sigma$ regions
as can be seen in the
right panel of Fig.~\ref{fig:123} and the
$g_\chi$ column
of Fig.~\ref{fig:corner}.

\begin{figure}[htbp]
\includegraphics[width=0.6\textwidth]{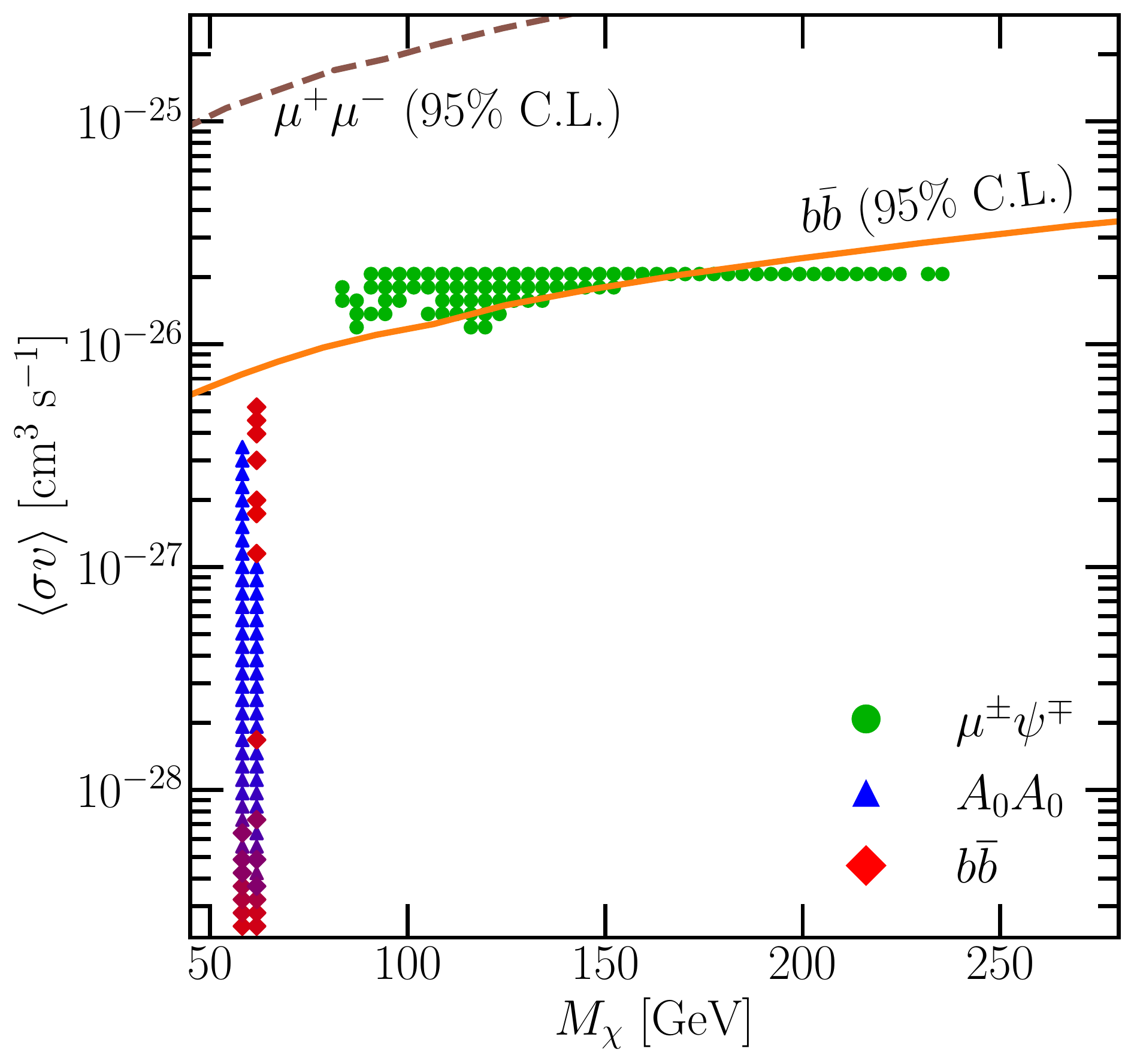} 
\caption{
The $2\sigma$ allowed samples projected to ($M_{\chi}$, $\langle\sigma v\rangle$) plane. 
The upper limit of dSphs gamma ray data for final state $b\overline{b}$~\cite{Oakes:2019ywx} and 
$\mu^+\mu^-$~\cite{Ackermann:2015zua} is depicted for reference. 
Different color and shape indicate the dominant DM annihilation channel.
} 
\label{fig:Mchi_sigmav}
\end{figure}

In Fig.~\ref{fig:Mchi_sigmav}, we project the samples which agree with all the constraints 
and the new measurement Eq.~\eqref{eq:E989} within $2\sigma$ to ($M_\chi$, $\sv$) plane. 
The cross section of the Higgs funnel is
mostly below the typical value
$3\times 10^{-26}$~cm$^3$s$^{-1}$ because the DM velocity in the present universe 
becomes $10^{-3}$ in units of the speed of light.
The resonance condition is no longer maintained. 
Except from the Higgs funnel, the annihilation cross section in rest of the regions is 
governed by $\chi\overline{\chi}\to \mu^\pm\psi^\mp$. 
The VLML can eventually decay to $b\overline{b}\mu^\pm$ and 
the final state of DM annihilation is $2b2\mu$.  
We therefore plot the $95\%$ upper limits from dSphs gamma ray data by final states
$b\overline{b}$~\cite{Oakes:2019ywx} (orange line) and 
$\mu^+\mu^-$~\cite{Ackermann:2015zua} (green dashed line) as comparison. 
At the mass range $100\gev \lesssim M_\chi \lesssim 200\gev$, 
the allowed parameter space may be further probed by 
the dSphs gamma ray data but a detailed analysis with correct 
gamma ray spectrum is needed.

\subsection{LHC}
\label{sec:LHC}

Before the end of this section, we comment some possible searches of this
model at the LHC\@.
Besides the Higgs boson invisible decay when $M_{\chi} < M_{H_0}/2$, a DM pair final state
can be explored through mono-X processes via the off-shell exchange of $H_0$, $A_0$~\cite{Kahlhoefer:2017dnp}.
In this model, the cross sections for mono-X processes are proportional to
$\sin 2\alpha$, making this exploration channel a challenging one at the LHC\@.

We then turn to the possible new spin-0 particle $A_0$.
Since $A_0\rightarrow b\overline{b}$ is the dominant decay mode,
the search for $H_0\to A_0 A_0\to b\overline{b}b\overline{b}$, as shown in Refs.~\cite{Aaboud:2018iil,Aad:2020rtv},
is crucial to confirm or rule out the excess presented in Ref.~\cite{ATLAS:2021ypo}. 
Other option is to use $pp\rightarrow ZA_0\rightarrow (l^{+}l^{-})(b\overline{b})$ to
confirm the existence of $A_0$ with $M_{A_0}=52$~GeV.
The production cross section is about $\sin^2\alpha\times 7.67$ pb. The well-known jet substructure techniques of Ref.~\cite{Butterworth:2008iy} can be applied to this search for $A_0\to b\overline{b}$.

Finally, even if the model is already constrained by the search of 
the pair production of VLML $\psi^{\pm}$ 
with multi-lepton signature as presented in the right panel of Fig.~\ref{fig:VLML_Xsec}, 
we can still explore multi-$b$ jets processes at the LHC in the near future. 
The signature for the single production of $\psi^{\pm}$ is $2\mu 2b$ and the possible SM backgrounds
are $t\overline{t}$, $t\overline{b}$, $b\overline{b}Z$.
Thanks to the larger cross sections of the $\psi^{\pm}$ pair production,
one can explore this channel by two signatures: $2b4\mu$ and $4b2\mu$.
The possible SM backgrounds for the former one are $t\overline{t}Z$ and $b\overline{b}ZZ$  
while the SM backgrounds for the later one are $t\overline{t}t\overline{t}$, $t\overline{t}b\overline{b}$, 
and $t\overline{t}t\overline{b}$.

\section{Conclusion and discussion}
\label{sec:conclusion}

\begin{figure}[htbp]
\begin{centering}
\includegraphics[width=0.8\textwidth]{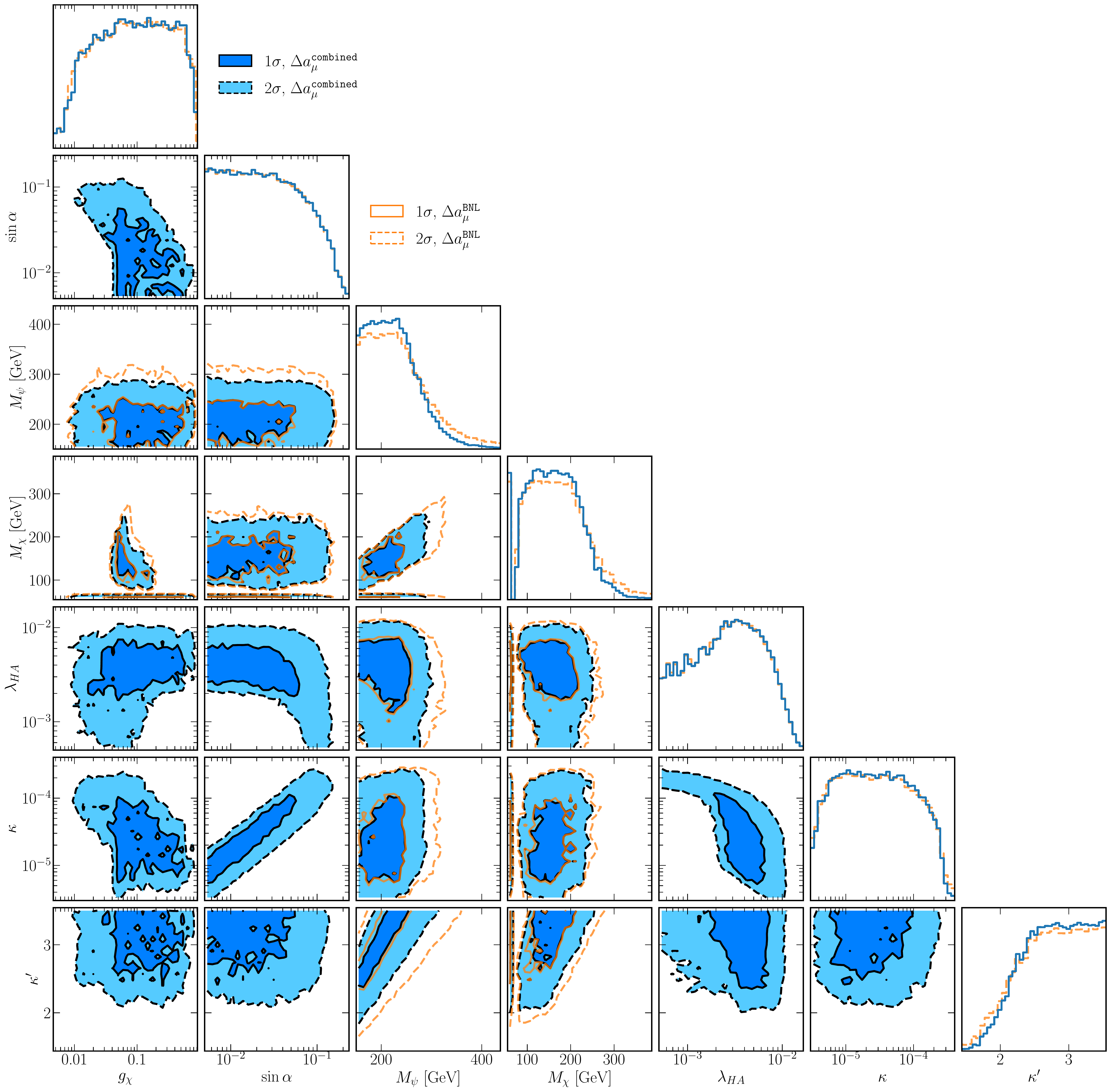} 
\caption{
The likelihood distribution among seven scan parameters:
$g_{\chi}$, $\sin\alpha$, $M_{\psi}$, $M_{\chi}$, $\lambda_{HA}$, $\kappa$,
$\kappa^{\prime}$ in our global analysis.
The 1$\sigma$ and 2$\sigma$ contours corresponding to using the $(g - 2)_\mu$
from the E821 experiment at BNL are shown in orange, only when the change is noticeable.
}
\label{fig:corner}
\end{centering}
\end{figure}

The simplified DM models are common approach for DM phenomenological studies.
DM and at least one mediator are two indispensable ingredients inside these models.
This opens up the possibility of discovering a mediator before finding the actual DM, thus helping
us narrow down the regions worth exploring and the possible interactions between DM and the SM\@.
Motivated by a local 3.3$\sigma$ deviation at $M_{A_0}=52$~GeV in the
$H_0\to A_0 A_0\to b\overline{b}\mu^{+}\mu^{-}$ search at ATLAS Run 2,
we proposed a model that this spin-0 particle $A_0$ is a pseudoscalar mediator.
Moreover, recently the Muon $g-2$ Collaboration at Fermilab hinted at BSM physics with
a reported 4.2$\sigma$ deviation from the SM in the combined $(g-2)_\mu$ measurements.
We found that a new vector-like muon lepton (VLML) can explain both
ATLAS Higgs boson exotic decay excess and $(g-2)_{\mu}$.
In this renormalizable DM model, we involve a Dirac fermion $\chi$ and
a pseudoscalar $A$, both SM singlets, plus an extra VLML $\psi^{\pm}$.  

We comprehensively constrain the parameter space from this model using
LHC Higgs boson data, DM measurements and electron and muon EDMs.
We found that due to the close relationship between $\kappa^{\prime}$, $M_{\psi^{\pm}}$ and $(g-2)_\mu$
these parameters are bounded as $\kappa^{\prime}>1.8$ and $M_{\psi}<315$~GeV at 2$\sigma$.
Both $\sin\alpha$ and $\lambda_{HA}$ are strongly affected by limits on
Higgs boson exotic and invisible decays
resulting in the upper bounds $\sin\alpha \lesssim 0.15$ and $\lambda_{HA} \lesssim 0.01$.
To have a prediction consistent with
${\rm BR}(H_0\rightarrow A_0 A_0\rightarrow b\overline{b}\mu^{+}\mu^{-})\sim 3.5\times 10^{-4}$
from ATLAS data, we found that $\kappa$ is positively correlated with $\sin\alpha$
as shown in Fig.~\ref{fig:Higgs}.
We determined the limit to which we can enhance ${\rm BR}(A_0\rightarrow\mu^{+}\mu^{-})$ with larger $\kappa$
considering that we need to increase $\sin\alpha$ as well to satisfy
$10^{-3} < {\rm BR}(A_0\rightarrow b\overline{b})\times {\rm BR}(A_0\rightarrow\mu^{+}\mu^{-}) < 4\times 10^{-2}$.
Note that $\kappa$ and $\sin\alpha$  are also constrained by
$H_0\rightarrow\mu^{+}\mu^{-}$, $H_0\rightarrow\gamma\gamma$
and electron and muon EDMs measurements, therefore, their values are quite restricted.  
In addition to the Higgs mediated resonance regions $\chi\overline{\chi}\to f\overline{f}$,
the major DM annihilation channels in our model are $\chi\overline{\chi}\to \mu^\pm\psi^\mp$
via $H_0/A_0$ exchanges and $\chi\overline{\chi}\to H_0 A_0$. 
Thanks to the pseudoscalar mediator, the constraints from DM direct detection can be safely ignored.

In summary, the renormalizable simplified DM model presented here simultaneously explains
the ATLAS Higgs boson exotic decay excess and the recently reported $(g-2)_{\mu}$ result. 
We summarize the 1,  2 and $3\sigma$ allowed regions of all the model parameters 
in the triangle plot Fig.~\ref{fig:corner}.
Moreover, we have proposed ways to further confirm the existence of $A_0$ with a mass $M_{A_0}=52$~GeV
and searches for VLML $\psi^{\pm}$ at the LHC\@.
Additionally, DM annihilation to $2\mu 2b$ is an interesting signature for indirect detection.
Here, the first muon is primary produced but the second muon 
together with a pair of b-quarks come from VLML decay. 
The raised either electron or gamma ray spectra can be very different 
with the conventional DM annihilation scenario whose two final state particles carry the same energy.  
We will return to this in a future work.

\begin{acknowledgments}
The analysis presented here was done using the resources of the
high-performance T3 Cluster at the Institute of Physics, Academia Sinica.
This  work is supported in part by KIAS Individual Grant, No.PG075301 (CTL) at Korea Institute for Advanced Study. 
Y.-L.~S.~Tsai was funded by the Ministry of Science and Technology Taiwan 
under Grant No. 109-2112-M-007-022-MY3.
The work of R.~Ramos is supported by the Ministry of Science and Technology of 
Taiwan under Grant No. 108-2811-M-001-550.
\end{acknowledgments}

\newpage

\appendix

\section{Electron and muon electric dipole moments} 
\label{sec:append}

In this appendix, we develop the contributions to electron and muon EDMs in our model.
The effective Lagrangian for the lepton $l$ EDM can be written as
\begin{equation}
{\cal L}^l_{\text{EDM}} = -\frac{i}{2}d^E_l F^{\mu\nu}\overline{l}\sigma_{\mu\nu}\gamma_5 l.
\end{equation} 
From the interactions in Eq.~\eqref{Eq:HA-int}, we determine the explicit contributions from two-loop Barr-Zee type diagrams~\cite{Ellis:2008zy,Giudice:2005rz,Li:2008kz} for both electron and muon EDMs
and display the resulting expressions in what follows.

\subsection{Two-loop Barr-Zee EDMs}

\begin{figure}[htbp]
\begin{centering}
\includegraphics[width=0.3\textwidth]{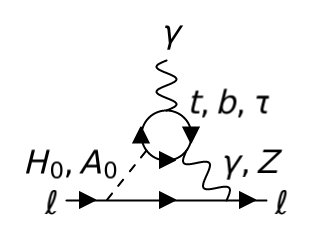} 
\includegraphics[width=0.3\textwidth]{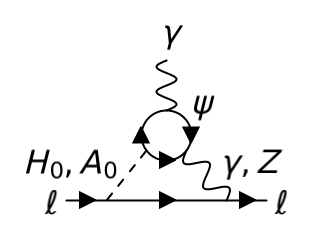}
\includegraphics[width=0.3\textwidth]{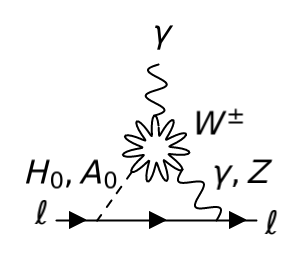}
\caption{
Two-loop Barr-Zee-type diagrams for electron and muon EDMs in our model.
Note that only the diagram in the middle contributes to electron EDM\@. 
}
\label{fig:Barr-Zee}
\end{centering}
\end{figure}

The electron EDM can be calculated from the two-loop Barr-Zee type diagram in the middle of Fig.~\ref{fig:Barr-Zee}.
The result can be written as 
\begin{equation}
d^E_e = (d^E_e)^{\gamma H_0/A_0} + (d^E_e)^{Z H_0/A_0}
\end{equation}
where 
\begin{equation}
\left(\frac{d^E_e}{e}\right)^{\gamma H_0/A_0} = \pm\frac{\alpha^2_{em}m_e}{8\pi^2 s^2_W m_W M_{\psi}}\sin 2\alpha\frac{y}{g}g_{\rm loop}(\tau_{\psi H_0/A_0})
\end{equation}
and
\begin{equation}
\left(\frac{d^E_e}{e}\right)^{Z H_0/A_0} = \pm\frac{\alpha^2_{em}m_e M_{\psi}(-\frac{1}{4}+s^2_W)}{32\pi^2 c^2_W s^2_W m_W M^2_{H_0/A_0}}\sin 2\alpha\frac{y}{g}\int^1_0 dx\frac{1}{x}J(\tau_{ZH_0/A_0}, \frac{\tau_{\psi H_0/A_0}}{x(1-x)})
\end{equation}
with $\tau_{xy}=M^2_x /M^2_y$.
The notation $H_0/A_0$ denotes summation of the contributions from $H_0$ and $A_0$
with the upper(lower) overall sign for $H_0$($A_0$).

Similarly, the muon EDM from all the two-loop Barr-Zee type diagrams in Fig.~\ref{fig:Barr-Zee} can be written as
\begin{equation}
d^E_{\mu} = (d^E_{\mu})^{\gamma H_0/A_0} + (d^E_{\mu})^{Z H_0/A_0}
\end{equation}
where 
\begin{align}
\left(\frac{d^E_{\mu}}{e}\right)^{\gamma H_0/A_0} = &
\pm\sum_{q = t,b}\frac{3\alpha^2_{em}Q^2_q}{4\sqrt{2}\pi^2 s^2_W M_{\psi}}\sin 2\alpha\frac{\kappa^{\prime}\kappa}{g^2}f_{\rm loop}(\tau_{qH_0/A_0}) 
\nonumber  \\ & 
\pm\frac{\alpha^2_{em}}{4\sqrt{2}\pi^2 s^2_W M_{\psi}}\sin 2\alpha\frac{\kappa^{\prime}\kappa}{g^2}f_{\rm loop}(\tau_{\tau H_0/A_0}) 
\nonumber  \\ &
\pm\frac{\alpha^2_{em}m_{\mu}}{8\pi^2 s^2_W m_W M_{\psi}}\sin 2\alpha\frac{y}{g}g_{\rm loop}(\tau_{\psi H_0/A_0})
\nonumber  \\ &
\mp\frac{\alpha^2_{em}}{16\sqrt{2}\pi^2 s^2_W M_{\psi}}\sin 2\alpha\frac{\kappa^{\prime}\kappa}{g^2} {\cal J}^{\gamma}_W (M_{H_0/A_0})
\end{align}
and 
\begin{align}
\left(\frac{d^E_{\mu}}{e}\right)^{Z H_0/A_0} = & 
\pm\frac{\alpha^2_{em} m^2_t (-\frac{1}{4}+s^2_W)(\frac{1}{4}-\frac{2}{3}s^2_W)}{32\sqrt{2}\pi^2 c^2_W s^4_W M_{\psi}M^2_{H_0/A_0}}\sin 2\alpha\frac{\kappa^{\prime}\kappa}{g^2} 
\int^1_0 dx\frac{1-x}{x}J(\tau_{ZH_0/A_0}, \frac{\tau_{tH_0/A_0}}{x(1-x)})
\nonumber  \\ & 
\mp\frac{\alpha^2_{em} m^2_b (-\frac{1}{4}+s^2_W)(-\frac{1}{4}+\frac{1}{3}s^2_W)}{64\sqrt{2}\pi^2 c^2_W s^4_W M_{\psi}M^2_{H_0/A_0}}\sin 2\alpha\frac{\kappa^{\prime}\kappa}{g^2} 
\int^1_0 dx\frac{1-x}{x}J(\tau_{ZH_0/A_0}, \frac{\tau_{bH_0/A_0}}{x(1-x)}) 
\nonumber  \\ & 
\mp\frac{\alpha^2_{em} m^2_{\tau} (-\frac{1}{4}+s^2_W)^2}{64\sqrt{2}\pi^2 c^2_W s^4_W M_{\psi}M^2_{H_0/A_0}}\sin 2\alpha\frac{\kappa^{\prime}\kappa}{g^2} 
\int^1_0 dx\frac{1-x}{x}J(\tau_{ZH_0/A_0}, \frac{\tau_{\tau H_0/A_0}}{x(1-x)}) 
\nonumber  \\ & 
\pm\frac{\alpha^2_{em}m_{\mu} M_{\psi}(-\frac{1}{4}+s^2_W)}{32\pi^2 c^2_W s^2_W m_W M^2_{H_0/A_0}}\sin 2\alpha\frac{y}{g}\int^1_0 dx\frac{1}{x}J(\tau_{ZH_0/A_0}, \frac{\tau_{\psi H_0/A_0}}{x(1-x)}) 
\nonumber  \\ & 
\pm\frac{\alpha^2_{em}(-\frac{1}{4}+s^2_W)}{16\sqrt{2}\pi^2 s^4_W M_{\psi}}\sin 2\alpha\frac{\kappa^{\prime}\kappa}{g^2} {\cal J}^Z_W (M_{H_0/A_0})
\end{align}
Again, the upper(lower) overall sign is for $H_0$($A_0$). 
The loop functions $f_{\rm loop}(\tau)$, $g_{\rm loop}(\tau)$, ${\cal J}^{G=\gamma ,Z}_W (M_i)$, and $J(a,b)$
can be found in Refs.~\cite{Ellis:2008zy,Ellis:2010xm,Abe:2013qla}.

\end{document}